\documentclass[useAMS,usenatbib]{mn2e}

\usepackage{graphicx}
\usepackage{lscape}

\begin{document}

\title[Infrared Spectroscopy of AGB and Post-AGB Stars]
{Infrared Spectroscopic Study of a Selection of AGB and Post-AGB Stars}
\author[V. Venkata Raman and B.G. Anandarao]{V. Venkata
Raman${^1}$\thanks{vvenkat@prl.res.in} and B.G. Anandarao${^1}$
\thanks{anand@prl.res.in} \\
$^1$Physical Research Laboratory (PRL),
Navrangpura, Ahmedabad - 380009, India}

\date{}

\pagerange{\pageref{firstpage}--\pageref{lastpage}} \pubyear{2007}

\maketitle

\label{firstpage}

\begin{abstract}

    We present here near-infrared spectroscopy in the H and K bands of a selection of
    nearly 80 stars that belong to various AGB types, namely S type, M type and SR type.
    This sample also includes 16 Post-AGB (PAGB) stars.
    {\bf From these spectra, we seek correlations between the equivalent widths 
    of some important spectral signatures and 
    the infrared colors that are indicative of mass loss. Repeated spectroscopic 
    observations were made on some PAGB stars to look for spectral variations.
    We also analyse archival SPITZER
    mid-infrared spectra on a few PAGB stars to identify spectral features due to
    PAH molecules providing confirmation of the advanced stage of their evolution. 
    Further, we model the SEDs of the stars (compiled from archival data) and compare
    circumstellar dust parameters and mass loss rates in different types.}

    {\bf Our near-infrared spectra show that in the case of M and S type stars, 
    the equivalent widths of the CO(3-0) band are moderately 
    correlated with infrared colors, suggesting a possible
    relationship with mass loss processes. A few PAGB stars revealed
    short term variability in their spectra, indicating episodic mass loss: the cooler stars showed
    in CO first overtone bands and the hotter ones showed in HI Brackett lines. 
    Our spectra on IRAS 19399+2312 suggest that it is a transition object. 
    From the SPITZER spectra, there seems to be a
    dependence between the spectral type of the PAGB stars and the strength of the PAH features.
    Modelling of SEDs showed among the M and PAGB stars that the higher the mass loss rates,
    the higher the [K-12] colour in our sample.}
\end{abstract}


\section{Introduction}

During their asymptotic giant branch (AGB) stage,
intermediate mass stars undergo substantial mass loss triggered by pulsation shocks
and radiation pressure (e.g., Willson \cite{will00}; van Winckel \cite{vanw03}).
Near-infrared spectroscopy is one of the recognized tools to
study the mass loss process (e.g., Kleinmann \& Hall \cite{klei86};
Volk, Kwok \& Hrivnak \cite{volk99};
Lancon \& Wood \cite{lanc00}; Bieging, Rieke \& Rieke \cite{beig02}; Winters et al. \cite{wint03}). The
near-infrared JHK bands, contain several important
diagnostic lines that can be used as signatures to probe the atmospheres of these stars
(Hinkle, Hall \& Ridgway \cite{hink82}). 

One of the most dramatic manifestations of the AGB stage is the high rate of mass loss
that is mainly attributed to pulsational levitation of atmosphere followed by expulsion
of matter by the radiation pressure (Wood \cite{wood79}; Bowen \cite{bowe88};
Gail \& Sedlmayr \cite{gail88}; Vassiliadis \& Wood \cite{vass93}; Habing \cite{habi96};
Feast \cite{feas00}; Winters et al. \cite{wint00}; Willson \cite{will00};
Woitke \cite{woit03}; Garcia-Lario \cite{garc06}).
One of the issues that we would address is the possible signatures of
pulsational mass loss in the spectra of AGB stars.
We would also like to study the infrared spectral signatures that the intermediate mass stars leave
while evolving beyond AGB stage to become Planetary Nebulae (e.g., Kwok \cite{kwok00})

In this work we report near-infrared (H and K bands) spectroscopy on a
selection of about 80 AGB stars of different types, namely M type, S type,
Semi-regular variables (SR) and post-AGB (PAGB) candidates (Section 2).
In Section 3.1, we try to seek significant differences among these types and
correlate spectral signature strengths with pulsation parameters or
mass loss indicators (such as the infrared color indices) in case of M
and S and SR types.

   \begin{figure*}
   \centering
\includegraphics[scale=0.8]{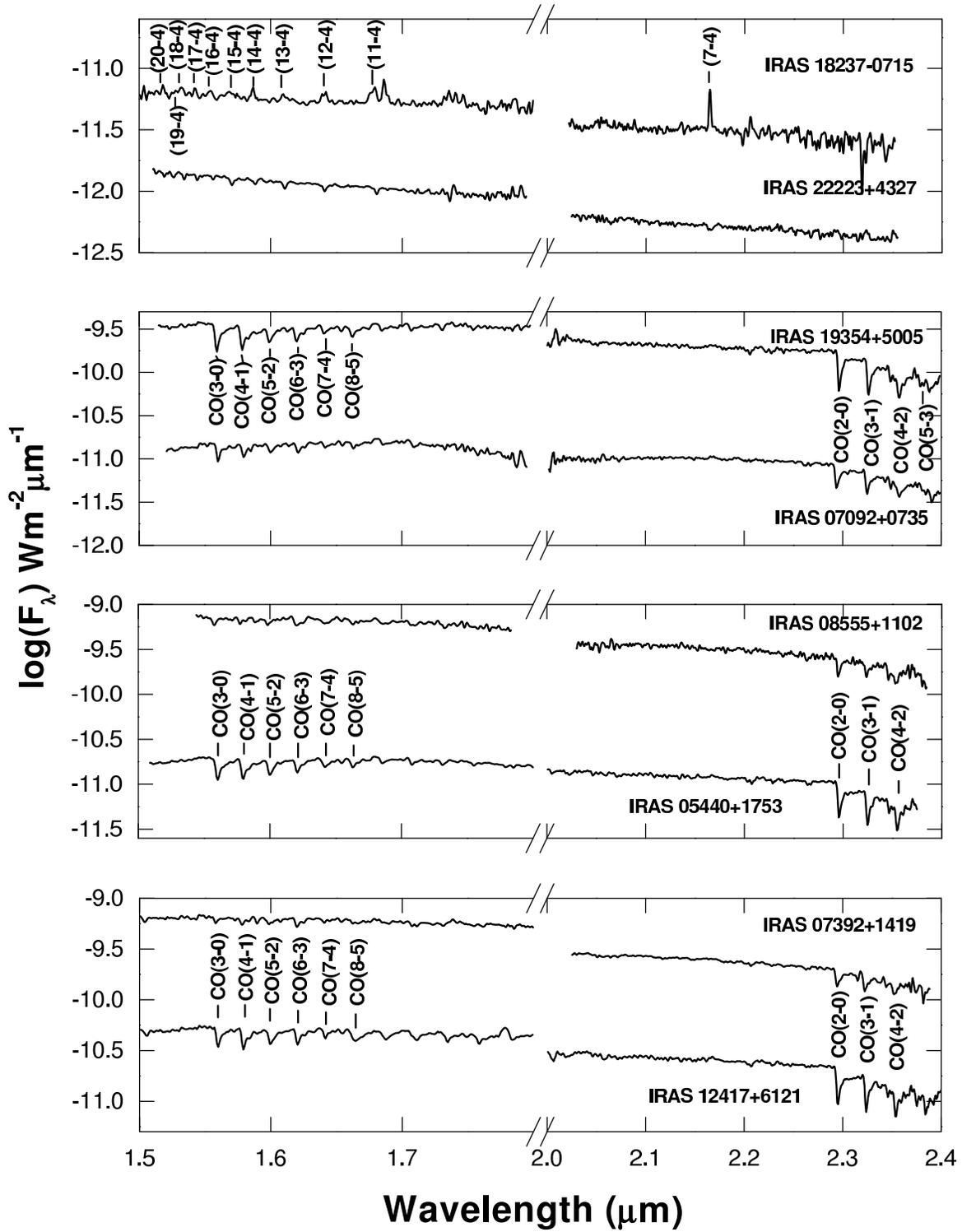}
   \caption{Mt Abu near-infrared spectra of a selection of AGB and PAGB stars.
   From top to bottom
   the panels show PAGB, M, S and SR type stars giving two examples per category.}
              \label{nirspec}%
    \end{figure*}

   \begin{figure*}
   \centering
  \includegraphics[scale=0.6,angle=-90]{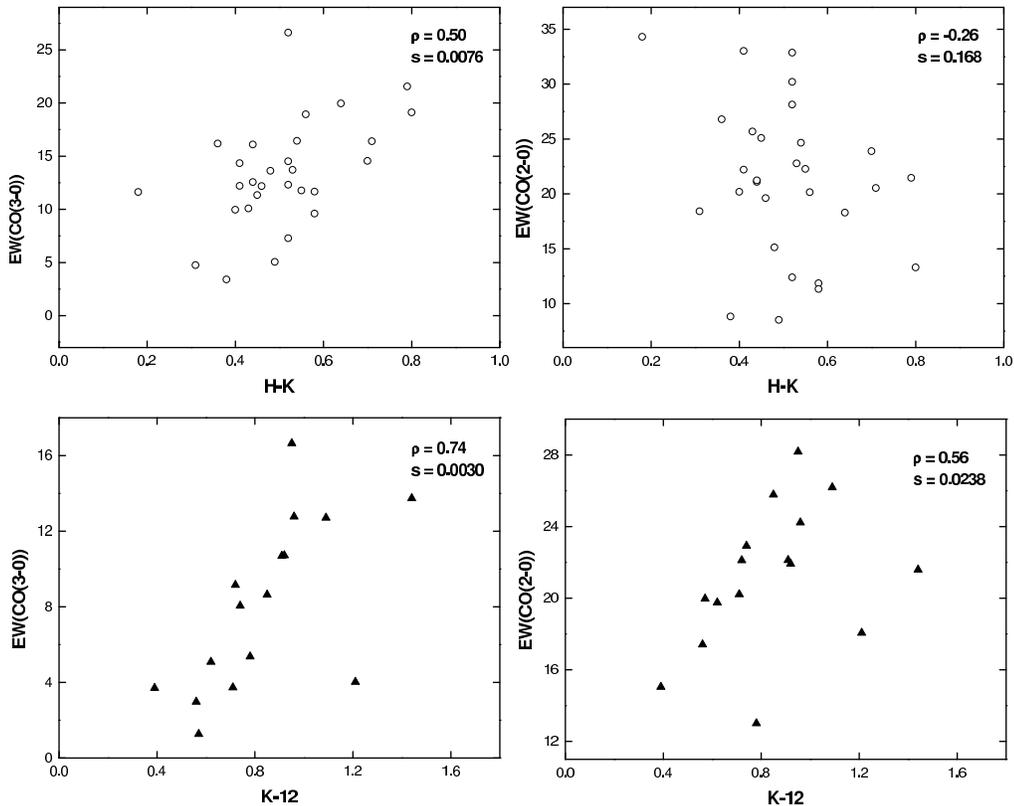}
   \caption{Plots of Equivalent Widths (in \AA) of the CO(3-0) line (left panels) and
   CO(2-0) line (right panels) with the IR colors [H-K] for M stars (shown as open circles in top panels)
    and [K-12] for S type stars (filled triangles in bottom panels). The Spearman rank correlation
    coefficient ($\rho$) and its significance (s) are given in the corresponding figures.}
              \label{co2030}%
    \end{figure*}

Further, we present the SPITZER archival spectra in the mid-infrared
region on a few post-AGB stars in order to find out the evolutionary
stage of these stars in terms of the spectral features seen in the spectra (Section 3.2).
Following this, using the DUSTY code, we model the infrared
spectral energy distributions of these sample stars, constructed from archival data
in the visible, near and far-infrared regions to investigate basic differences in the photospheric and
circumstellar parameters among the various types of AGB stars (Section 3.3).
Some interesting aspects on variation of spectral lines in a few PAGB stars are discussed in Section 4.
Section 5 lists important conclusions of the present work.

\begin{center}
\begin{table*}
\caption{2MASS \& IRAS Colors and Mt Abu Equivalent Widths (in \AA) of the Spectral Features of the Sample M Stars}
\label{mstar1}
\centering
\begin{tabular}{l c c c c c c c c c c r}
\hline\hline
Star &  Sp. Type  & $\phi$ & H-K & K-12 & CO(3-0) & CO(4-1) & CO(H)   &    CO(2-0)  &  CO(3-1)& CO(K)
 & $\dot{M}$    \\
\hline
02234-0024 &  M4e   & 0.34 & 0.44  &  2.31  & 16.1 &  7.8 & 51.5 &  21.1 &  6.9 &  61.9  &  $2.6(-6)^{\dagger}$ \\
03082+1436 &  M5.5e & 0.33 & 0.45  &  1.99  & 11.3 &  9.2 & 64.2 &  25.1 & 13.3 &   62.3 &  $9.8(-7)^{\dagger}$\\
03489-0131 &  M4III & ...  & 0.31  &  1.86  &  4.8 &  5.3 & 61.9 &  18.4 & 20.2 &  45.9  &  1.3(-6)\\
04352+6602 &  Sev   & 0.44 & 0.50  &  1.22  &  ... &  ... &  ... &  31.8 & 29.8 & 101.2  &  $1.4(-7)^{\dagger}$ \\
05265-0443 &  M7e   & 0.57 & 0.49  &  1.32  & 13.6 & 10.1 & 53.1 &  15.1 &  7.8 &  22.9  &  $4.6(-7)^{\dagger}$ \\
05495+1547 &  S7.5  & 0.49 & 0.59  &  2.79  & 11.6 &  8.4 & 47.4 &  11.9 & 16.0 &  47.6  &  3.5(-6) \\
06468-1342 &  S     & ...  & 0.52  &  1.03  & 14.5 & 16.6 & 85.0 &  32.9 & 31.2 & 110.6  &  ...\\
06571+5524 &  Se    & 0.59 & 0.64  &  1.67  & 20.0 & 22.1 &105.6 &  18.3 & 25.6 & 52.8   &  $1.7(-6)^{\dagger}$ \\
07043+2246 &  S6.9e & 0.36 & 0.18  &  1.17  & 11.6 & 13.9 & 77.0 &  34.3 & 28.8 &  98.8  &  $3.1(-7)^{\dagger}$ \\
07092+0735 &  Se    & 0.30 & 0.71  &  3.02  & 16.4 & 15.4 & 72.1 &  20.5 & 25.8 &  81.0  &  2.9(-6) \\
07197-1451 &  S     & 0.26 & 0.55  &  2.32  & 11.8 & 8.0  & 65.0 &  22.3 &  9.7 &  43.5  &  4.6(-6) \\
07299+0825 &  M7e   & 0.57 & 0.36  &  1.90  & 16.2 & 10.5 & 64.1 &  26.8 & 14.1 &  65.3  &  $8.5(-7)^{\dagger}$ \\
07462+2351 &  SeV   & 0.96 & 0.52  & -0.90  & 12.3 & 13.8 & 68.8 &  28.1 & 33.9 & 108.0  &  ... \\
07537+3118 &        & 0.09 & 0.53  &  1.40  & 13.7 &  9.9 & 76.5 &  22.8 & 25.0 &  77.5  &  ...\\
07584-2051 &  S2,4e & 0.19 & 0.44  &  1.00  & 12.6 & 11.0 & 78.4 &  21.2 & 17.9 &  39.1  &  ... \\
08188+1726 &  SeV   & ...  & 0.41  &  2.28  & 14.3 & 13.6 & 66.1 &  22.2 & 24.8 &  71.3  &  ... \\
08308-1748 &  S     & ...  & 0.43  &  0.72  & 10.1 & 15.9 & 72.6 &  25.7 & 19.1 &  92.9  &  1.7(-7) \\
08416-3220 &        & ...  & 0.41  &  0.97  & 12.2 & 14.2 & 75.8 &  33.0 & 35.0 & 105.3  &  ...\\
08430-2548 &        & ...  & 0.46  &  1.51  & 12.2 & 13.0 & 68.3 &  19.6 & 16.6 & 80.2   &  ...\\
09411-1820 &  Me    & 0.31 & 0.49  &  1.63  &  5.1 & 7.3  & 33.2 &   8.5 &  5.7 &  22.1  &  $7.9(-7)^{\dagger}$ \\
12372-2623 &  M4    & 0.92 & 0.40  &  1.39  &  9.9 & 14.8 & 64.2 &  20.2 & 23.8 &  50.8  &  ... \\
17001-3651 &  M     & 0.70 & 0.58  &  2.17  &  9.6 &  8.9 & 57.0 &  11.3 &  8.9 &  35.8  &  $1.9(-6)^{\dagger}$ \\
17186-2914 &  S     & ...  & 0.70  &  1.76  & 14.5 & 13.2 & 80.7 &  23.9 & 26.1 &  89.2  &  2.9(-6) \\
17490-3502 &  M     & 0.60 & 0.54  &  2.26  & 16.4 & 16.1 & 78.1 &  24.6 & 17.6 &  95.8  &  ...\\
17521-2907 &  Se    & 0.11 & 0.56  &  1.64  & 18.9 & 19.5 & 88.4 &  20.1 & 21.9 &  71.8  &  4.6(-6)\\
18508-1415 &  SeV   & 0.44 & 0.38  &  1.00  &  3.4 &  5.1 & 64.3 &  8.8  &  5.5 &   31.1 &  ...\\
19111+2555 &  S     & 0.57 & 0.80  &  4.36  & 19.1 & 17.6 & 66.8 &  13.3 & 11.2 &  24.5  &  $4.6(-6)^{\dagger}$ \\
19126-0708 &  S..   & 0.90 & 0.79  &  3.81  & 21.6 & 19.1 &102.0 &  21.5 & 23.5 &  59.4  &  $2.9(-6)^{\dagger}$ \\
19354+5005 &  S..   & 0.24 & 0.52  &  2.28  & 26.6 & 22.8 &108.0 &  30.2 & 33.8 &  98.2  &  $1.5(-6)^{\dagger}$ \\
23041+1016 &  M7e   & 0.87 & 0.52  &  2.40  &  7.3 &  9.0 & 50.7 &  12.4 &  6.3 &  18.7  &  $1.5(-6)^{\dagger}$\\
\hline
\end{tabular}
\end{table*}

\begin{table*}
\caption{2MASS \& IRAS Colors and Mt Abu Equivalent Widths (in \AA) of Spectral Features of the Sample S Stars}
\label{sstar1}
\centering
\begin{tabular}{l c c c c c c c c c r}
\hline\hline
Star &  Sp. Type  & H-K  & K-12  & CO(3-0) & CO(4-1) & CO(H)     &  CO(2-0)  &  CO(3-1)& CO(K)
 & $\dot{M}$   \\
\hline
04599+1514 &  M0       &  0.49  &  0.96  & 12.8 & 17.3 & 80.8 &  24.2 & 28.6 & 99.4 &  $1.4(-7)^{\dagger}$\\
05036+1222 &           &  0.60  &  1.09  & 12.7 & 15.1 & 80.9 &  26.2 & 28.1 & 54.3 &  ...\\
05374+3153 &  M2Iab    &  0.16  &  1.44  & 13.7 & 12.2 & 77.5 &  21.6 & 26.6 & 88.3 &  $7.3(-7)^{\dagger}$  \\
07247-1137 &  Sv..     &  0.34  &  0.85  &  8.6 & 10.7 & 57.8 &  25.8 & 19.1 & 74.5 &  ... \\
07392+1419 &  M3II-III &  0.09  & -1.44  &  5.3 &  8.4 & 38.9 &  16.7 & 14.7 & 55.3 &  7.3(-7)  \\
09070-2847 &  S        &  0.25  &  0.91  & 10.7 & 12.8 & 74.1 &  22.1 & 37.1 & 59.2 &  ...\\
09152-3023 &           &  0.47  &  1.21  &  4.0 &  9.4 & 44.2 &  18.1 & 14.4 & 39.1 &  ... \\
10538-1033 &  M...     &  0.29  &  0.74  &  8.1 &  4.5 & 44.6 &  22.9 &  9.6 & 64.9 &  ... \\
11046+6838 &  M...     &  0.40  &  0.39  &  3.7 &  9.5 & 41.1 &  15.0 & 10.4 & 54.7 &  ... \\
12219-2802 &  S        &  0.30  &  0.72  &  9.2 &  9.4 & 57.0 &  22.1 & 17.8 & 69.8 &  ...\\
12417+6121 &  SeV      &  0.41  &  0.95  & 16.7 & 18.0 & 82.4 &  28.2 & 29.5 & 92.2 &  $1.0(-7)^{\dagger}$ \\
13421-0316 &  K5       &  0.22  &  0.57  &  ... &  8.3 & 40.9 &  20.0 & 14.9 & 59.8 &  ...\\
13494-0313 &  M...     &  0.42  &  0.56  &  3.0 &  9.2 & 55.5 &  17.4 & 20.0 & 47.4 &  ...\\
14251-0251 &  S...     &  0.34  &  0.71  &  3.7 &  9.1 & 45.8 &  20.2 & 16.7 & 60.0 &  ...\\
15494-2312 &  S        &  0.38  &  0.62  &  5.1 &  8.1 & 46.1 &  19.8 & 17.2 & 59.3 &  ...\\
16205+5659 &  M...     &  0.18  &  0.78  &  5.4 &  7.1 & 45.3 &  13.0 & 19.1 & 61.6 &  ...\\
16209-2808 &  S        &  0.54  &  0.92  & 10.7 & 13.6 & 73.9 &  21.9 & 27.2 & 79.5 &  1.0(-7)\\
\hline
\end{tabular}
\end{table*}

\begin{table*}
\caption{2MASS \& IRAS Colors and Mt Abu Equivalent Widths (in \AA) of Spectral Features of the Sample SR Stars}
\label{srstar1}
\centering
\begin{tabular}{l c c c c c c c c c c r}
\hline\hline
Star &  Sp. Type  & $\phi$ & H-K & K-12 & CO(3-0) & CO(4-1) & CO(H)     &  CO(2-0)  &  CO(3-1)& CO(K)
 & $\dot{M}$   \\
\hline
04030+2435 &  S4.2V    & ...  & 0.40  &   0.63  &  4.1 &   8.3  &  45.0  &	 16.2 &   17.9  &   34.1  &  ...\\
04483+2826 &  CII      & ...  & 0.38  &   1.04  &  7.0 &  10.9  &  60.3  &	 24.5 &   17.6  &   58.3  &  1.4(-7)\\
05213+0615 &  M4       & 0.41 & 0.39  &   1.37  &  9.3 &  11.1  &  64.8  &	 19.3 &   21.4  &   56.9  &  ...\\
05440+1753 &  S0.V     & 0.33 & 0.66  &   1.89  & 22.7 &  19.5  &  99.5  &	 33.2 &   30.3  &   88.1  &  1.4(-6)\\
06333-0520 &  M6       & 0.85 & 0.38  &   1.20  & 19.7 &   10.1 &   77.8 &	 19.1 &    24.8 & 61.5    &  1.7(-7)\\
08272-0609 &  M6e      & ...  & 0.40  &   1.47  & 11.3 &   7.8  &  51.7  &	 19.1 &   15.1  &    81.4 &  5.6(-7)\\
08372-0924 &  M5II     & ...  & 0.36  &   1.57  &  5.5 &   6.5  &  52.3  &	 18.5 &   12.4  &   49.0  &  1.7(-7)\\
08555+1102 &  M5III    & ...  & 0.26  &   1.28  &  7.4 &   7.9  &   61.8 &	 15.5 &    13.9 &   43.7  &  2.4(-7)\\
09076+3110 &  M6IIIaSe & ...  & 0.31  &   1.20  & 13.5 &  11.6  &  65.9  &	 18.5 &    18.8 &   56.7  &  $3.8(-7)^{\dagger}$\\
10436-3459 &  S        & ...  & 0.37  &   1.61  & 10.5 &  15.4  &  61.2  &	 30.3 &   23.9  &   79.0  &  $9.5(-7)^{\dagger}$\\
15492+4837 &  M6S      & 0.77 & 0.40  &   1.58  & 12.7 &   13.0 &  63.8  &	19.9  &   24.5  &   44.4  &  $9.5(-7)^{\dagger}$\\
16334-3107 &  S4.7:.V  & ...  & 0.43  &   1.13  & 15.0 &  22.9  &  93.4  &	 22.5 &   29.7  &  104.8  &  $1.0(-7)^{\dagger}$\\
17206-2826 &  M..      & ...  & 0.42  &   0.82  & 10.3 &  11.4  &  68.1  &	 15.2 &   19.6  &  63.8   &  $1.0(-7)^{\dagger}$\\
17390+0645 &  SV...    & ...  & 0.36  &   0.92  & 12.1 &   7.6  &  56.7  &	 29.8 &   14.2  &  70.9   &  ...\\
19497+4327 &           & ...  & 0.60  &   0.82  &  6.7 &  11.5  &  59.2  &	 25.2 &   18.1  &  73.8   &  ...\\ 
\hline
\end{tabular}
\end{table*}

\begin{table*}
\caption{2MASS \& IRAS Colors and Mt Abu Equivalent Widths (in \AA) of Spectral Features of the Sample PAGB Stars}
\label{pagb1}
\centering
\begin{tabular}{l c c c c c c c c r}
\hline\hline
Star   &  Sp. Type &  H-K & K-12 & Pa$\beta$ & Br$\gamma$ & CO(K) & 21$\mu$m & $T_{d}$ & $\dot{M}$\\
\hline
Z02229+6208& G9a     & 0.44  & 6.38  & ...   &  9.2  &  8.4  & ...   & 300      & 2.9(-5)\\
04296+3429 & G0Ia    & 0.49  & 7.34  &  ...  &  4.1  &  17.7 & -2.11 & 270      & $4.5(-5)^{\dagger}$\\
05113+1347 & G8Ia    & 0.25  & 5.98  &  ...  &  6.5  &  12.9 & ...   & 350      & 2.1(-5) \\
06556+1623 & Bpe     & 1.11  & 5.10  &  ...  & -52.4 &  ...  & ...   & 1200/150 & 1.0(-5)\\
07134+1005 & F5Ia    & 0.10  & 6.45  &  ...  &  5.1  &  ...  & -1.73 & 280      & $2.3(-5)^{\dagger}$\\
07284-0940 & K0Ibpv  & 0.23  & 5.64  &  ...  &  ...  &  10.9 & ...   & 450      & 1.0(-5) \\
07430+1115 & G5Ia    & 0.44  & 6.35  &  ...  &  3.9  &  28.9 & ...   & 320      & 3.3(-5) \\
17423-1755 & Be      & 1.48  & 5.30  &  ...  &  -5.6 & -23.6 & ...   & 125      & $4.9(-6)^{\dagger}$ \\
18237-0715 & Be      & 0.55  & 2.29  & -25.6 & -14.9 &  ...  & ...   & 70	    & 5.8(-6) \\
19114+0002 & G5Ia    & 0.27  & 4.84  & ...   & 5.4   & ...   & ...   & 120      & $4.9(-5)^{\dagger}$\\
19157-0247 & B1III   & 1.19  & 5.76  & ...   &  ...  & -10.8 & ...   &  270     & $1.1(-6)^{\dagger}$\\
19399+2312 & B1e     & 0.11  & 4.80  & -6.2  & -15.6 &  ...  & ...   & ...      & ...\\
22223+4327 & F9Ia    & 0.24  & 4.44  &  7.5  &  5.8  &  ...  & -1.31 & 125      & $1.0(-5)^{\dagger}$\\
22272+5435 & G5Ia    & 0.38  & 5.55  &  4.4  &  3.4  &  30.2 & -0.83 & 250      & $1.6(-5)^{\dagger}$ \\
22327-1731 & A0III   & 0.90  & 4.95  &  3.8  &  ...  &  28.7 & ...   & 800      & 5.0(-6)\\
23304+6147 & G2Ia    & 0.36  & 6.48  &  3.4  &  5.1  &  12.8 & -2.53 & 250      & $3.5(-5)^{\dagger}$\\
\hline
\end{tabular}
\end{table*}
\end{center}

\section{Near-Infrared Spectroscopic Observations:} Near-infrared
spectroscopic observations were made in the J(only for some of the sample PAGB stars), H and K bands
on the selection of stars, using the NICMOS-3 near-infrared camera/spectrograph
at the Cassegrain focus of the 1.2 m telescope at Mt. Abu, Western India, during the
period 2003-07. The observations were made at a spectral resolution of $\sim$ 1000.
Spectrophotometric standard stars of A0V type were used
for each programme star at similar air mass to remove the telluric absorptions
and for relative flux calibration
(cf. Lancon \& Rocca-Volmerange \cite{lanc92}). For sky background subtraction,
two traces were taken for each star for each cycle of integration at two spatially separated
positions along the slit within 30 arcsec. Typical overall integration times varied
between 0.6 to 160 sec for the brightest and
faintest stars in our sample (in each of the three bands). The atmospheric OH vibration-rotation
lines were used to calibrate wavelength. The HI absorption lines from the
standard star spectra are removed by interpolation. The resulting spectra of the
standard star were used for the removal of telluric lines from the program star by
ratioing. The ratioed
spectra were then multiplied by a blackbody spectrum at the effective temperature
corresponding to the standard star to obtain the final spectra. Typically the S/N ratios
at the strongest and the weakest spectral features varied between 100 and 3 respectively.
Data processing was done using standard spectroscopic tasks (e.g., APALL) inside IRAF.

   \begin{figure*}
   \centering
\includegraphics[angle=-90,scale=0.7]{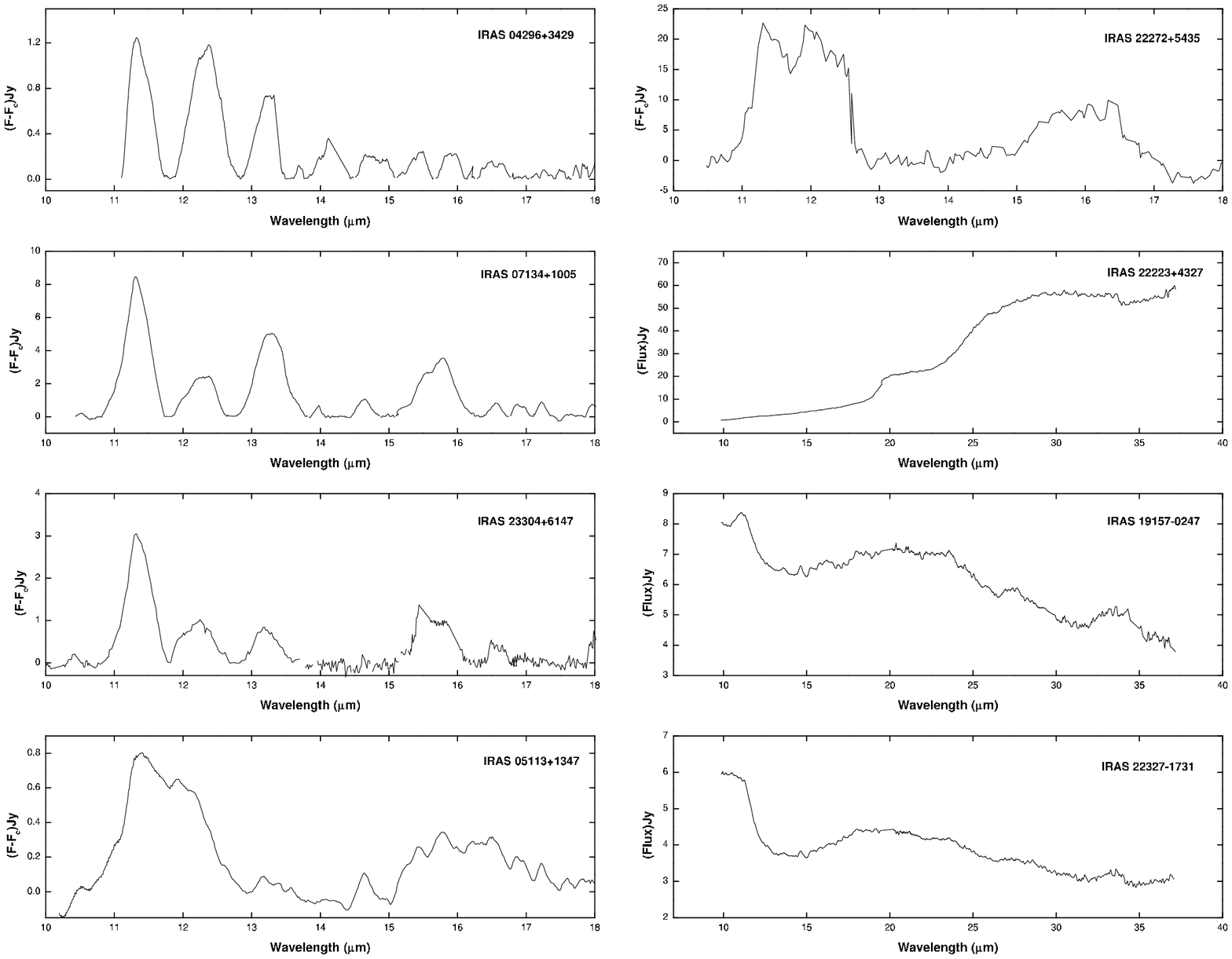}
   \caption{SPITZER Spectra  for some P-AGB Stars showing mid-infrared
   PAH features in emission (for all the four spectra
   shown in the left-side panels and the one in the top right panel,
   continuum was subtracted using (fifth order) polynomial fitting in segments).
   The spectra for IRAS 22272+5435 were obtained by re-analysing the ISO data and shown here for
   comparison.}
              \label{spitsp1}%
    \end{figure*}

     \begin{figure*}
   \centering
\includegraphics[scale=0.5]{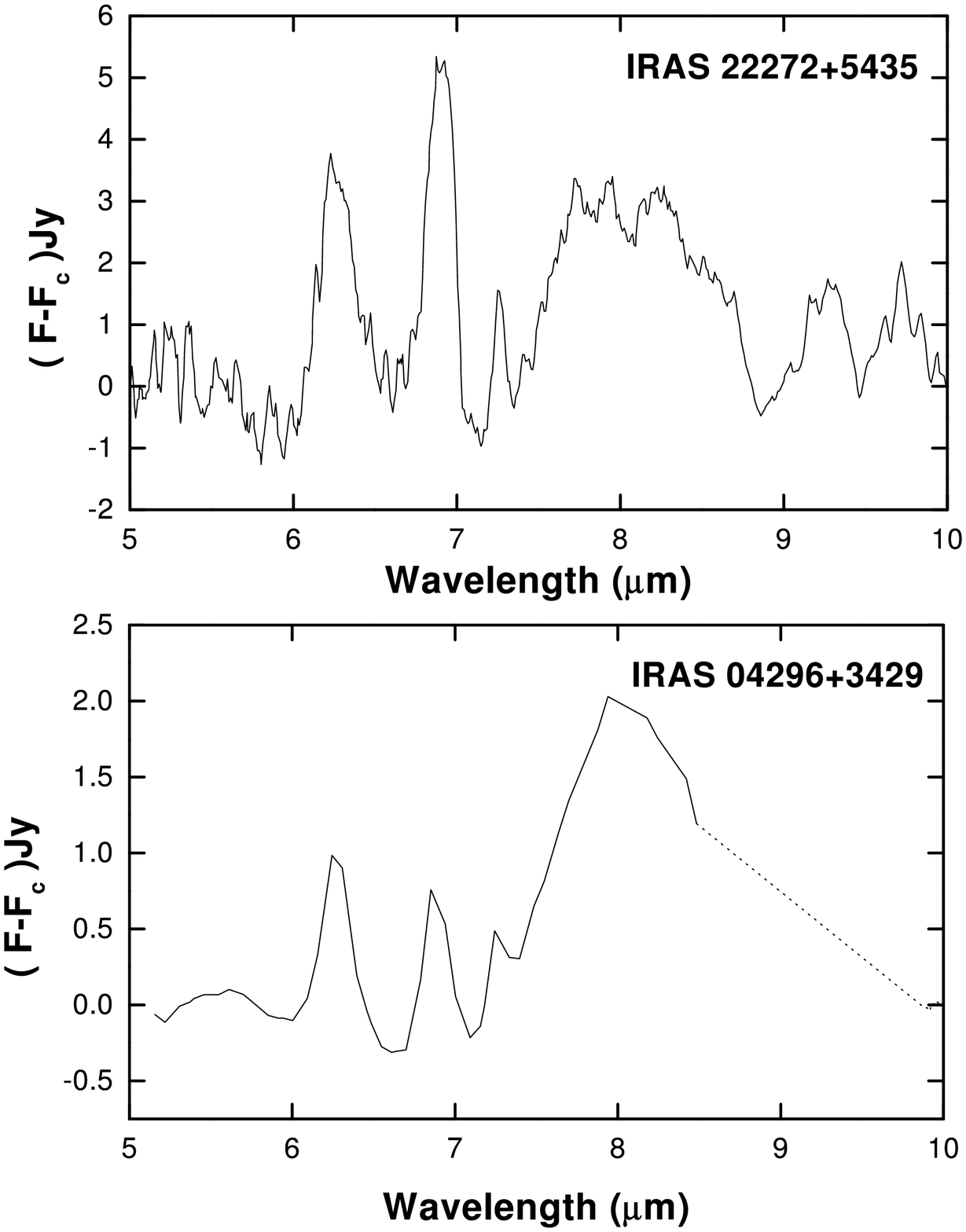}
   \caption{SPITZER Spectra of IRAS 04296+3429 showing mid-infrared
   (in the range 6-10 $\mu$m) PAH features in emission (continuum-subtracted as in Fig 3).
   The spectra for IRAS 22272+5435 were obtained by re-analysing the ISO data and is shown here for
   comparison.}
              \label{spitsp2}%
    \end{figure*}


\subsection{The Sample Stars:}  The programme stars were selected from published
literature/catalogues (Fouque et al. \cite{fouq92}, Kerschbaum \& Hron \cite{kers94},
Chen et al. \cite{chen95}, Wang \& Chen \cite{wang02}, and Stasinska et al. \cite{stas06}).
Our sample contains about 80 stars
with the selection based on the observability at Mt. Abu during the period November-May
and JHK band magnitudes brighter than 8-9. Among these stars, 30 are M type;
17 S type; 15 SR type and 16 are PAGB stars (some of which are known to be
transition/proto-PNe, see Ueta et al. \cite{ueta03} and Kelly \& Hrivnak \cite{kell05}).
The greater bias towards M types in the sample is due to the
fact that they are brighter in K band than the others. Tables 1-4 list all the stars
in the four categories along with the [H-K] and [K-12] colors 
computed from 2MASS and IRAS photometric archival data. The phases ($\phi$) of the variable
stars of M and SR types, corresponding to the dates of our observations, are also listed in the tables 
(epochs taken from Kholopov et al. \cite{khol88}).
The spectral types are taken from SIMBAD data base.

\section{Results and Discussion:}
Fig 1 shows some typical spectra for different types considered here. In general,
we find in the AGB stars, the photospheric absorption lines of Na I doublet at 2.21 $\mu$m,
Ca I triplet at $\sim$ 2.26 $\mu$m,
as well as Mg I at 1.708 $\mu$m in all the sample stars. Brackett series lines
of HI were found mostly among the PAGB stars where a hotter central source is
believed to be present. In addition, the CO vibration-rotation lines of first and second
overtone bands ($\Delta v = 2$ and $\Delta v = 3$ respectively) were seen
in a large number of M, S and SR type stars and in a few PAGB stars.
Some PAGB stars showed HI Brackett emission lines 
(IRAS 06556+1623, 17423-1755, 18237-0715 and 19399+2312 for the 
last one of which our spectra are new) and an indication of fainter
emission in CO. This could be due to the fact that as the central star hots up, the
CO lines become progressively weaker
and finally disappear when CO dissociates (at 11.2 eV). Around this time the HI recombination
lines start becoming prominent in emission. Our spectra on IRAS 06556+1623 and 17423-1755
are quite comparable to those of Garcia-Hernandez et al. \cite{garc02}, with a faint 
detection of H${_2}$ S(1) 1-0 line at 2.12 $\mu$m.
 For the PAGB object IRAS 22327-1731, our spectra showed for the first time absorption 
in HI lines and CO bands. Some of the PAGB objects that showed variability are discussed in section 4. 
Equivalent widths (EWs) were
computed for all the spectral features that are detected in our sample stars. The
errors in the equivalent widths are mainly from the S/N ratio of the lines. To that
extent the errors are estimated to be about 3 \AA, mostly applicable to
the low S/N ($\sim$ 3) line detections, such as the metallic lines; 
but for high S/N ($\geq$ 5) line/bands the errors are $\leq$ 1 \AA.
Tables 1-4 list the EWs (in \AA, positive for absorption and negative for emission) of the
CO(3-0, 4-1) second overtone bands in H band, CO(2-0, 3-1) first overtone bands in K band,
and HI Brackett series lines (only for PAGB stars), for the programme stars of each category.

\subsection{Correlations and their implications:}

Based on physical processes, very interesting correlations were found among spectral features
observed in AGB stars (Kleinmann \& Hall \cite{klei86}; Lancon \& Wood \cite{lanc00};
Bieging, Rieke \& Rieke \cite{beig02}). We tried to see if our sample too shows these. 
Kleinmann \& Hall \cite{klei86} and Bieging, Rieke \& Rieke \cite{beig02} 
have found that the CO(2-0) and CO(3-1) band head
strengths are correlated. From the tables 1-3, we can see this trend; in addition, 
one can see a correlation between CO(3-0) \& CO(4-1) bands as well.

   \begin{figure*}
   \centering
 \includegraphics[scale=0.7,angle=-90]{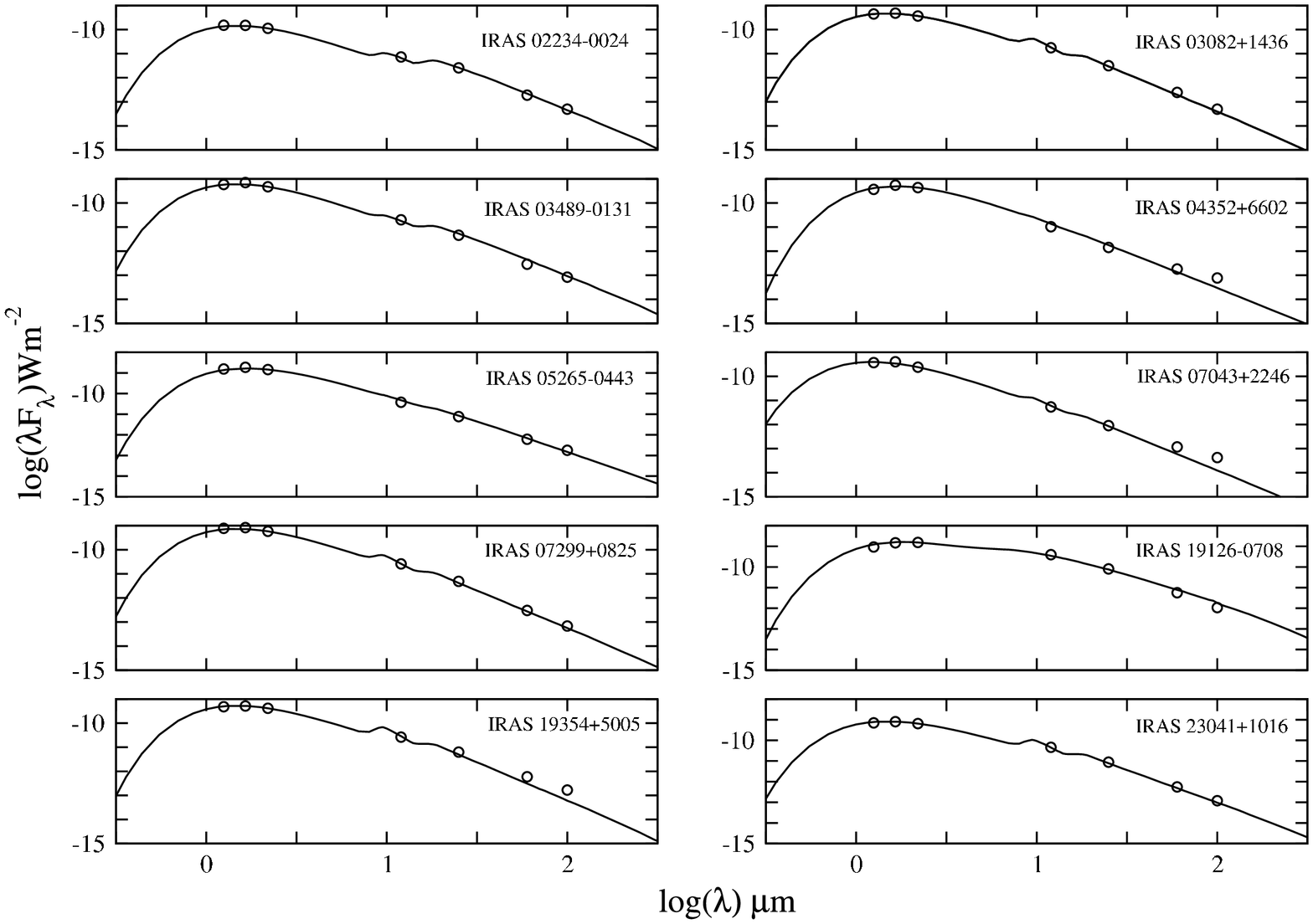}
   \caption{Model spectral energy distributions (full lines) of some M type stars compared with observed data from
   literature (open circles). See text for model parameters
    and explanation.}
              \label{models1}%
   \end{figure*}

   \begin{figure*}
   \centering
 \includegraphics[scale=0.7,angle=-90]{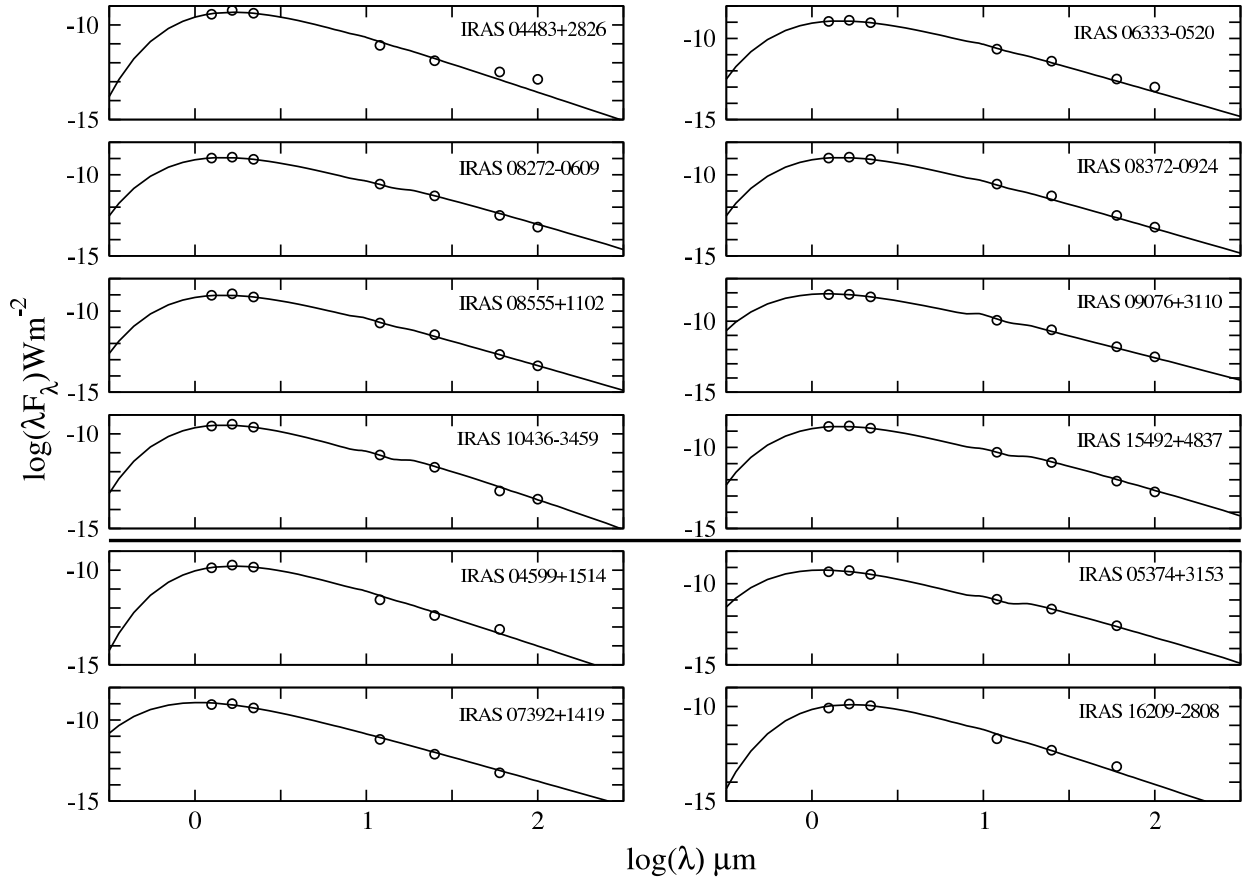}
   \caption{Model spectral energy distributions (full lines) of some SR type (in the top 8 panels)
   and S type stars (in the bottom 4 panels below the horizontal line) compared with observed data from
   literature (open circles). See text for model parameters
    and explanation.}
              \label{models1}%
   \end{figure*}

   \begin{figure*}
   \centering
 \includegraphics[scale=0.7,angle=-90]{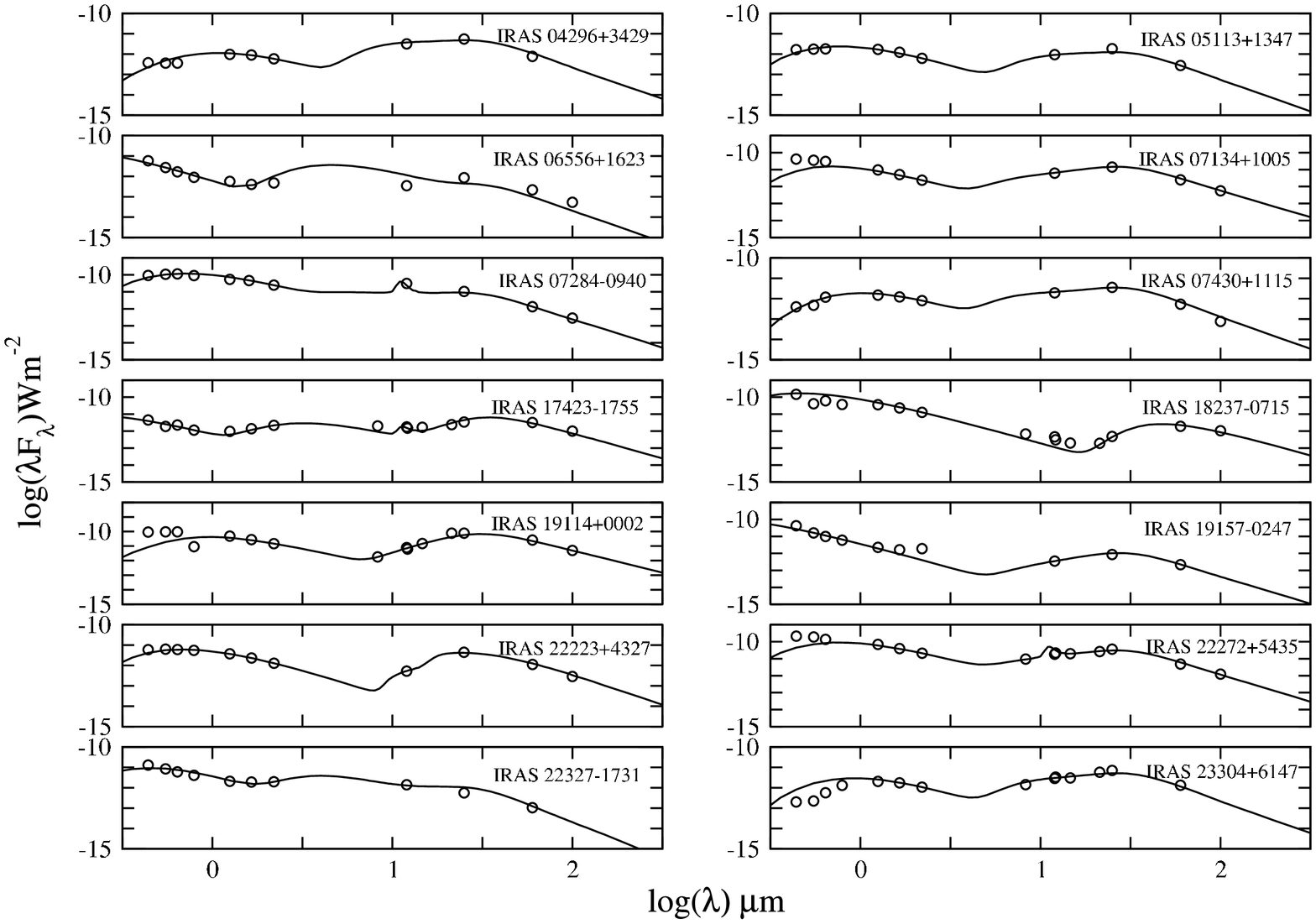}
   \caption{Model spectral energy distributions (full lines) of some PAGB stars compared with observed data from
   literature (open circles). See text for model parameters
    and explanation.}
              \label{models1}%
   \end{figure*}


The first overtone bands of CO ($\Delta v = 2$) in the K band
arise at T $\sim$ 800 K partly in the photosphere and
partly in the circumstellar envelopes; while the second overtone bands
($\Delta v = 3$) in the H band arise at T $\sim$ 3000-4000 K entirely
in the photospheric layers (e.g., Hinkle, Hall \& Ridgway \cite{hink82}; Emerson \cite{emer96}).
As a result, the pulsational effects may be more pronounced in the
$\Delta v = 3$ bands than in the $\Delta v = 2$ bands; while the
former will reflect the circumstellar matter properties better.
Thus we would expect a better correlation between EW of
$\Delta v = 3$ bands with pulsation period P than the $\Delta v = 2$ bands.
It should be noted here that for the CO bands the S/N ratio is high(in many cases $\geq$ 5) and
hence the errors in EW are $\leq$ 1 \AA.
Fig 2  shows equivalent widths of CO(3-0) and CO(2-0) plotted against
 the color [H-K]  for M type stars. The phases of these
variable stars are not necessarily the same between our observations and 2MASS or
IRAS data acquisition (see Tables 1 for phases during our observations).
Also shown in Fig 2  are the EWs of the CO lines from the S type stars
(against the color [K-12]). The trend of correlation for CO(3-0) band
is quite clear. For SR stars, we did not find any such trend.

In order to quantify the degree of correlation between two parameters,
we have used the Spearman's rank correlation method that does not assume any functional
relationship between them (e.g., used by Loidl, Lancon \& Jorgensen \cite{loid01}).
Fig 2 shows only those correlations with CO(3-0) 
for which the rank coefficient  ($\rho$) $\geq$ 0.50 and 
its significance ($s$) $\leq$ 0.01 (i.e., a chance occurrence of 1 in 100).  
We find that for M stars, for the CO(3-0) vs [H-K], $\rho$) is 0.50 with $s$ of
0.008 (a chance occurrence of 8 in 1000); while for S stars, for CO(3-0) vs [K-12] $\rho$ is 0.74
with $s$ of 0.003. These
are moderately significant correlations as may also be visually seen from the plots. 
In comparison, the coefficients and their significance are quite poor for the CO(2-0) plots in Fig 2.
For M stars, one can also see a trend in the
CO ($\Delta v = 3$) vs log P better than CO($\Delta v = 2$) vs log P(not shown).
Usually the mass loss is correlated well with the pulsation
period; as also by the color indices [H-K] and [K-12](Whitelock et al. \cite{whit94}).
While the trends in Fig 2 are certainly beyond the errors in EW,
we find that the scatter is rather significant in the correlation plot between
the CO(3-0) band EWs and [H-K] and other mass loss indicators (not shown) such as log P
and [K-12], although it is much less in comparison to the plots for
CO(2-0) band (only [H-K] is shown). 
Lancon \& Wood \cite{lanc00} and Loidl, Lancon \& Jorgensen \cite{loid01} argued 
that in the case of carbon-rich AGB stars, the CO lines arise deeper in
side the atmospheres (close to the photospheres) and hence may not
respond to dynamical effects such as pulsations as much as those lines that arise in the outer
atmospheres. In our sample M stars, a large number of them are oxygen-rich and hence
the pulsational phase effects may be significant and can possibly account for the scatter
in our plots.
   
   \begin{figure*}
   \centering
 \includegraphics[scale=0.6,angle=-90]{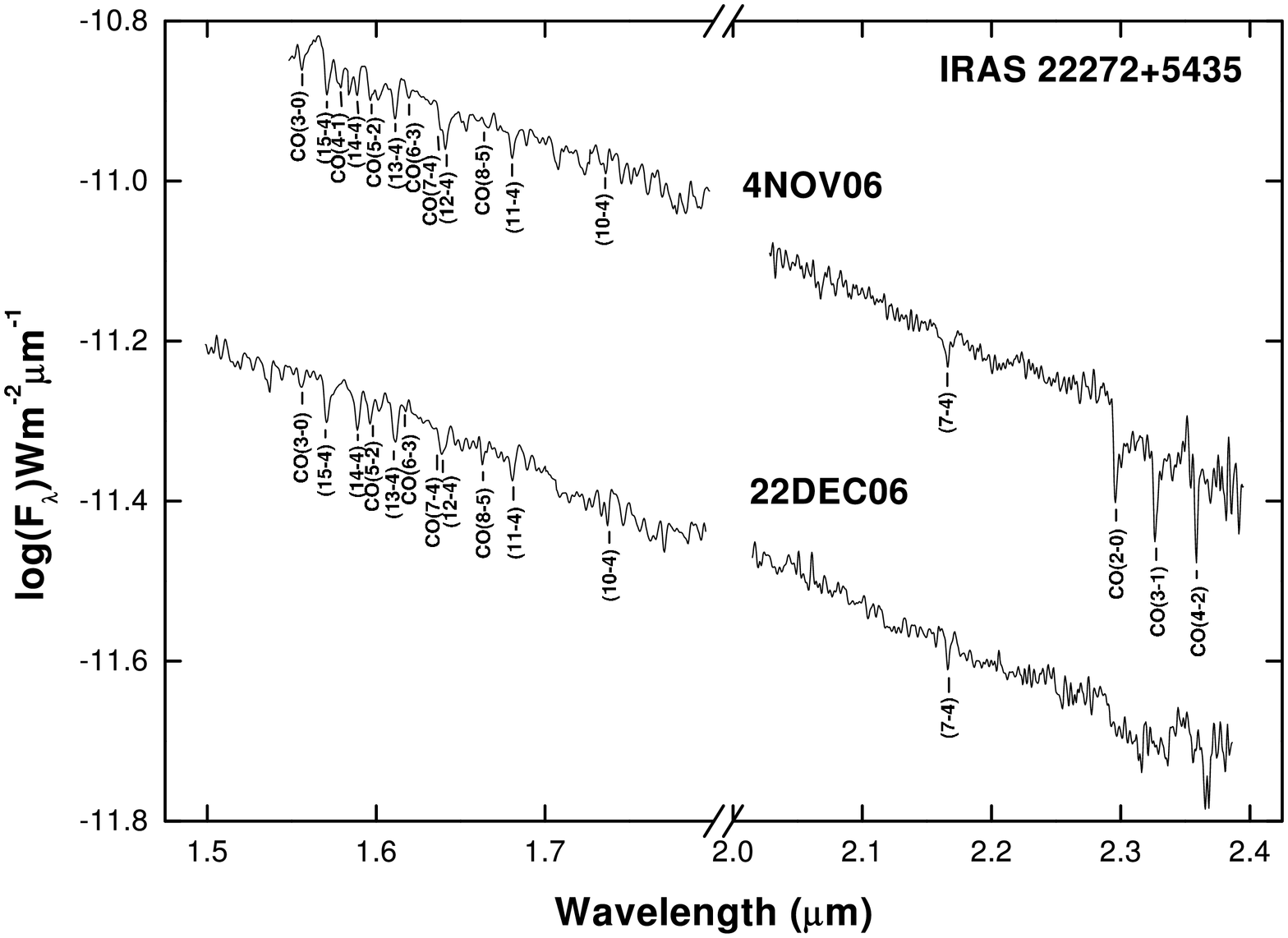}
   \caption{H and K band Spectra of the PAGB star IRAS 22272+5435 taken on two dates from Mt Abu.}
              \label{var22272}%
   \end{figure*}

\subsection{SPITZER Spectra on a few Post-AGB objects - PAH and 21 $\mu$m features:}
We have searched the SPITZER archival spectroscopic data in the mid-infrared region of
6-30 $\mu$m and succeeded in obtaining 7 PAGB stars for which the data were
available and unpublished and are not in the list of sources presented originally
by Hrivnak, Volk \& Kwok \cite{hriv00}.
The raw spectra (post BCD Data) were analyzed using the Spitzer
IRS Custom Extraction (SPICE) task under SPITZER data
processing softwares. Interesting spectra in the wavelength range of 10-18 $\mu$m
from these objects are shown in Fig 3 (where in addition to the 7 SPITZER stars,
we have shown one more PAGB star, namely IRAS 22272+5435, re-analysed from the ISO spectral archives).
Of these 8 stars, IRAS 19157-0247 and IRAS 22327-1731 are of early spectral types
and showed nearly featureless continua (see Fig 3).
For the spectra of those stars that showed the in-plane and out-of-plane bending
modes of C-H bonds in PAH molecules (e.g., Kwok \cite{kwok04} and Tielens \cite{tiel05}),
we have subtracted the continuum by (fifth order) polynomial fitting
in segments, in order to bring out the comparative strengths of the features.
Fig 4 shows the 6-10 $\mu$m spectra on two stars
for which the data are available; these stars show the C-C stretching modes of 6.2 and 6.9 $\mu$m
and the blended features around 8 $\mu$m due to C-C stretching and C-H in-plane bending modes.
A comparison of the spectra in the region of 10-18 $\mu$m common to about 6 PAGB objects
suggests that the intensity of the spectral features due to PAH molecules depends on the
spectral type of the star - being weakest or shallowest for cooler stars than for relatively hotter
stars. This can possibly be due to the UV radiation flux and the hardness that is available
to excite the molecules (Tielens \cite{tiel05}).

We have computed the ratio of the EWs of the F$_{7.9}$/F$_{11.3}$ feature
which is known to indicate whether the molecule is neutral or ionized (see Tielens \cite{tiel05}).
We find that the PAGB star IRAS 04296+3429, having a ratio of 11.3,
probably shows indications for ionized PAH molecules, while in IRAS 22272+5435,
with a ratio of 0.8, neutrals may be predominant (see Figs 3 and 4).
Further, the relative structure of PAHs may be inferred
from the ratio of F$_{11.3}$/F$_{12.7}$: a large ratio
indicates large compact PAHs, while a small ratio results when these molecules break up
into smaller irregular structures (Hony et al. \cite{hony01};
Sloan et al. \cite{sloa05}; Peeters et al. \cite{peet02}; Tielens \cite{tiel05}).
In our sample PAGB stars, we find (see Fig 3) that the objects
IRAS 07134+1005 and 23304+5147 have ratios of 4.7 and 3.6 respectively
and hence may have large compact PAH molecules
compared to IRAS 04296+3429 that has a ratio of 0.8. In IRAS 05113+1347 and
22272+5435 the SPITZER spectra show wide features at 7-8 and 11-12 $\mu$m regions. Such
broad features are usually attributed to more complex structures in PAH molecules
(see Kwok \cite{kwok04}).

Since the PAH modes arise by absorption of UV photons,
their presence indicates the onset of substantial UV flux from the central star and
hence the onset of transition/proto-PN phase. Mt. Abu spectra taken on all these
objects do not show the HI Brackett series in emission; they are
seen in absorption. Thus it is possible that the UV radiation is soft ($ \leq $ 13.6 eV)
as H ionization has not yet started in these objects.

Only three of our sample PAGB stars have SPITZER archival data beyond 18 $\mu$m, 
and all showed the 21 $\mu$m feature (Kwok, Volk \& Hrivnak \cite{kwok89,kwok99}; 
Volk, Kwok \& Hrivnak \cite{volk99} and Decin et al. \cite{deci98}), which was attributed
to TiC (von Helden et al. \cite{vonh00}) or SiC (Speck \& Hofmeister \cite{spec04}).
Since these objects are already known to have the 21 $\mu$m feature (Volk, Kwok \& Hrivnak \cite{volk99}), 
in Table 4 we give only the EWs of the feature (in $\mu$m), along with 
the EW for IRAS 07134+1005 computed from ISO archival data.
All the properties of the observed feature (the peak wavelength, width and red-side asymmetry)
resemble the SiC feature as shown by Speck \& Hofmeister \cite{spec04}).

\subsection{Modeling of Infrared Spectral Energy Distribution:}   Using the DUSTY
code (Ivezic \& Elitzur \cite{ivez97}), we have modeled the spectral energy
distributions (SED) in the
spectral region of 0.5 - 100 $\mu$m. 
This code incorporates full dynamical calculation for radiatively driven winds
in AGB stars. 
The photometric data on all the programme objects
were compiled mainly from 2MASS and IRAS archives. 
We have modelled the SEDs of all the programme stars except those that have 
upper-limits (quality = 1) in IRAS data in more than one band. 
For a majority of the programme stars IRAS LRS data are not available. 
Even in the ISO archives we could get only a few stars.  
Therefore, we have modelled only the photometric fluxes, assuming 
silicate dust for M, S and SR types; 
while for most of the PAGB stars, we required carbonaceous dust.
The dust grain data were taken from in-built tables in DUSTY. 
In a few cases we obtained data from DENIS
and MSX archives as well. In the model that assumes spherical geometry, we fix stellar parameters
like the photometric temperature (based on the visible and the near-IR data), inner dust
shell temperature, type of dust and size distribution (MRN assumed) and opacity at
0.5 $\mu$m (usually taken value between 0.1 and 1.0).
We have used density distribution relevant for
modelling the radiatively driven winds in AGB stars as incorporated in to DUSTY
by its authors. There would be 10-20 \% uncertainties in all the model output parameters.
We have generated nearly 150 models
using several combinations of the above parameters. We then take the fluxes computed by
the code and normalize with the observed 2 $\mu$m flux from each object and
construct the model SEDs. A $\chi^2$ test is performed to select the best fit model.
In general the model shows that the
M types have lower effective temperatures than the S and SR types. M types show
higher opacities by a factor of 2, than the S and SR types; S type being lowest.
However the inner dust shell temperature for all the types were found to be in the range
450-700 K. The SEDs of PAGB stars showed, in general, double-humped but flatter trend
than their AGB counterparts, indicating detached circumstellar shells (see e.g.,
Hrivnak \& Kwok \cite{hriv99}). Since some of these are transition objects their
spectral classes are much earlier than M types. The inner dust
shell temperatures for most of the PAGB stars, as listed in Table 4,
are found to be in the range of $\sim$ 100-300 K, significantly cooler than the values for AGB stars.
A warmer dust shell was required in addition to the cooler one in one of the 
PAGB stars (1200 K for IRAS 06556+1623). 
The dust parameters that we obtained are in good agreement with those in the published literature.  
A few examples of SEDs of M, S, SR and PAGB stars and their model fits are shown in Figs 5-7,
respectively.
The DUSTY AGB star model estimates mass loss rate also but with larger uncertainty of 30 \%.
The mass loss rates (in M$_{\odot}$/yr) thus obtained are listed in Tables 1-4 for all the types; 
the sign $\dagger$ shows stars for which the rates are independently available 
from CO rotational line observations by several authors, namely,
Loup, Forveille \& Omont \cite{loup93}, Sahai \& Liechti \cite{saha95}, 
Groenewegen \& de Jong \cite{groe98},
Jorissen \& Knapp \cite{jori98}, Winters et al. \cite{wint00}, Ramstedt et al. \cite{rams06},
for M, S and SR stars; and Woodsworth, Kwok \& Chan \cite{wood90}, Likkel et al. \cite{likk91},
Omont et al. \cite{omon93}, Hrivnak \& Kwok \cite{hriv99}, Bujarrabal et al. \cite{buja01},
Hoogzaad et al. \cite{hoog02} and Hrivnak \& Bieging \cite{hriv05} for PAGB stars. 
Our model mass loss rates are in reasonably
good agreement with those determined from CO rotational lines, for all cases.
From Tables 1 and 4, one can see a clear trend of increase of mass loss rate with increase of the color
[K-12] (e.g., Bieging, Rieke \& Rieke \cite{beig02}). {\bf Basically most of our sample M type
stars are in the ascending phase of their AGB stage and hence the mass loss rate
increases with the [K-12] index (see Whitelock et al. \cite{whit94} and Habing \cite{habi96}); while
the PAGB stars in our sample are already in the PPN stage, having just
passed the superwind mass loss event. Following this phase the mass loss would decrease.
In fact one can see (Table 4) that the mass loss rates for hot PAGB stars are distinctly less than those 
for cooler ones, possibly indicating the decrease of mass loss after a superwind event.}

\section{Spectral variations in some PAGB objects}

Most of the PAGB stars in our sample are identified as Proto-PNe (see Ueta et al. \cite{ueta03}
 and references therein).
Repeated spectroscopic observations were made on a few selected stars
IRAS 17423-1755, IRAS 18237-0715, IRAS 19399+2312 and IRAS 22272+5435 (see Table 4 for their spectral types).
Our results on several sources are compared with those
taken earlier by Hrivnak, Kwok \& Geballe \cite{hriv94},
Oudmeijer et al. \cite{oudm95} and recently by Kelly \& Hrivnak \cite{kell05}.
Here we report spectral variations in the CO
first overtone bands in IRAS 22272+5435; and in HI lines in stars of early
spectral type (hot PAGB stars).
In both cases the time scales of variability are short and may be associated with episodic mass loss.

\subsection{Variability of CO first overtone bands in IRAS 22272+5435:}

The spectra of this object of spectral type G5Ia show signatures 
that are probably indicative of its advanced stage of evolution
in the PAGB phase. It was shown by Hrivnak, Kwok \& Geballe \cite{hriv94} that the object
shows episodes of mass loss that was evident from month-scale variability in its spectra.
We obtained its spectra in H and K bands on two occasions separated by about 47 days
(4 Nov 2006 and 22 Dec 2006). EWs for 4 Nov 06 are given in Table 4.
Fig 8 shows the spectra and it is clear that the CO first
overtone bands in the K band seen on 4 Nov were absent on 22 Dec. But the Brackett HI absorption
lines were seen in both H and K band spectra on both the observations. Interestingly the H band
second overtone bands of CO were {\it still present} in the spectra on 22 Dec. This shows that
the mass loss reached the outer regions of the atmosphere where the temperature is about a few
hundreds of degrees K. The pulsations must be responsible for heating the matter in these regions
resulting in the excitation of these lines and filling the absorption seen earlier.

We have found variability of spectra in a few other PAGB stars similar to IRAS 22272+5435.
The H and K band spectra of IRAS 05113+1347, IRAS 19114+0002 and
IRAS 22223+4327 were observed by Hrivnak, Kwok \& Geballe \cite{hriv94}. Our
observations, made about 12 years later, show clear indication of
variability of CO in these stars which might have occurred during the intervening period.
In all these cases, the variations were found only in CO first overtone bands
while the HI lines did not change. {Oudmeijer et al. \cite{oudm95} found
CO second overtone bands in emission in the object IRAS 19114+0002 while our observations
show near absence of these bands which can be attributed to mass loss variations.}

   \begin{figure*}
   \centering
 \includegraphics[scale=0.6,angle=-90]{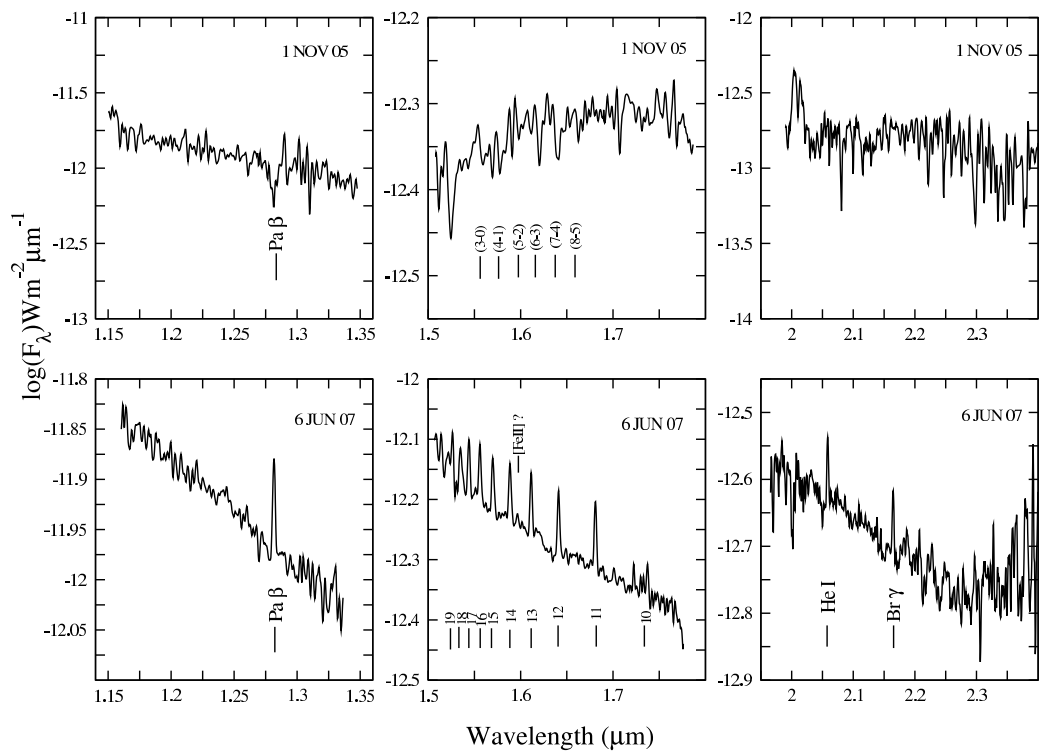}
   \caption{J, H and K band spectra of the PAGB star IRAS 19399+2312 taken on two dates from Mt Abu.}
              \label{var19399}%
   \end{figure*}

\subsection{Variability of HI lines in IRAS 19399+2312:}

On the basis of the IRAS colors, Gauba et al. \cite{gaub03} suggested that IRAS 19399+2312
(also known as V450 Vul) is a hot PAGB star possibly in transition to become a proto-PN.
It was classified as a B1III star by Parthasarathy et al. \cite{part00} but as B1IIIe by
Kohoutek \& Wehmeyer \cite{koho99}. 
Recently, Greaves \cite{grea04} argued that being a possible member of the cluster NGC 6823,
this star is unlikely to be a PAGB transition object but could be
a spectroscopic type Be star.
We made JHK band spectrometry at Mt Abu Observatory on this object on three occasions: first time
on 1 Nov 2005 and then recently on 15 May 2007 and again on 5-6 Jun 2007 (the EWs
obtained on 6 Jun 07 are given in Table 4). The 1 Nov 2005 spectra
showed some what shallow absorption in HI Paschen $\beta$ and Brackett series lines. The later spectra
however showed very prominent emissions in the HI lines, as well as a few permitted lines in HeI
and even forbidden lines in [FeII] (1.60 $\mu$m) at relatively lower S/N (see Fig 9).
We also notice at a rather low S/N but still discernable CO ($\Delta$v=3) lines
in absorption. The appearance of the CO absorption lines is not usual in Be stars, as also the
forbidden lines of metals ([FeII]). In view of the appearance of forbidden
lines (which was not reported earlier), and faint but significant CO lines,
we believe that this object may as well be a transition object. To our knowledge,
our observations are the first to show spectral variability in this object.
From the ratios of the Brackett series lines which deviate from the Case B assumption
(see for example, Lynch et al. \cite{lync00}), we infer variations in electron density
or opacity in the atmosphere of this object associated with the HI line ratio variations
that occurred between 15 May 2007 and 5/6 June 2007. 

We made repeated observations on two more PAGB stars of early spectral type:
for IRAS 17423-1755 (Be) on 29 May 03 and 13 May 07 and for
IRAS 18237-0715 (Be) on 28 May 04, 3 Nov 06 and 14 May 07.
In Table 4, the EWs for IRAS 17423-1755 are for the date 29 May 03; and for IRAS 18237-0715
the EWs are for 3 Nov 06.
We find variations in the intensities of HI Brackett series emission lines in both these objects.
In the case of IRAS 18237-0715 the Brackett line ratios matched well with the
Case B assumption on 28 May 04 and 3 Nov 06; while on 14 May 07 the ratios deviated from Case B.
Earlier in a detailed study Miroshnichenko et al. \cite{miro05} found spectral line variations in this object.
This star did not show CO lines (Table 4).
In the case of IRAS 17423-1755 on both the days of our observations the ratios deviated from Case B.
These observatons indicate the changes in electron density or opacity. This star showed CO first overtone lines
in emission (Table 4).
While Gauba et al. \cite{gaub03} found emission lines
in the optical spectra of IRAS 17423-1755 and 18237-0715 and hence termed these as transition objects,
the nature of the latter object is ambiguous and could be a luminous blue variable
according to the study of Miroshnichenko et al. \cite{miro05}. Our IR spectra of this object closely
resemble those of Miroshnichenko et al. \cite{miro05} and show P-Cygni type profiles with
double peaks (see Fig 1 top panel). Since we do not find CO lines in the spectra of this object
it is likely that this object can not possibly be a PAGB/transition object; or CO must have been dissociated.

In the case of the object IRAS 22327-1731 (type A0III) Oudmeijer et al. \cite{oudm95} found that the CO bands
as well as the HI lines were absent; our results (Table 4) show that the object displays the CO bands and
HI Paschen $\beta$ and Brackett $\gamma$ in absorption, thus showing variability.

\section{Conclusions:}

 \begin{enumerate}

  \item {\bf The equivalent widths of CO (3-0) show
  a trend of correlation with [H-K] in M type; while in S stars
  the CO (3-0) shows a positive correlation with [K-12].}

  \item {\bf The Mt Abu H and K band spectra also showed marked differences
  in the three types.} 

  \item {\bf Archival SPITZER spectra of a few PAGB stars showing PAH emission features
confirm their advanced stage of evolution into PPNe. There appears to be a dependence
of the strength of the PAH features on the spectral type.}

  \item {\bf Modeling of SEDs of a number of the programme stars showed
    marked differences in various generic types. 
    Model mass loss rates for M and PAGB stars show an increasing
    trend with the color [K-12] in our sample stars.}
    
  \item {\bf Our observations showed that several PAGB stars undergo short term spectral variability that
  is indicative of on-going episodic mass loss. Cooler PAGB stars showed variation in CO first overtone lines.
  In contrast, the hot PAGB stars showed variation in HI lines. The hot PAGB star IRAS 19399+2312 
  is identified as a transition object.}       

 \end{enumerate}

\section*{acknowledgements}
   The authors sincerely thank the anonymous referee for critical comments 
   that led to substantial improvement over the original version.  
   This research was supported by the Department of Space, Government of India.
   We thank the Mt Abu Observatory staff for help in observations with NICMOS. 
   We thank the SPITZER archival facilities. We acknowledge the DUSTY team for making 
   available the code for the astronomy community.

\end{document}